\documentstyle[12pt,epsf]{article}
\begin{document}
\pagestyle{empty}

\hfill{\sf Hot Points in Astrophysics}

\hfill{\sf JINR, Dubna, Russia, August 22-26, 2000}

\vspace*{2.cm}

\centerline{\large\bf Gamma-Ray Bursts: Astronomical Constraints}

\vspace*{0.2cm}

\begin{center}

Konstantin Postnov\\

{\it Sternberg Astronomical Institute, Moscow State University, 119899
Moscow, Russia} 

\end{center}

\vspace*{0.2cm}

\begin{abstract}

After presenting a short history of gamma-ray burst (GRB) studies, 
we discuss the current constraints on GRB models which follows
from astronomical observations. We concentrate on the energetics 
of the GRBs with known redshifts and the association of 
the well-localized GRBs with star forming regions in remote 
galaxies. We also discuss implications of the observed GRB rate. 
Arguments are given favoring possible connection of GRBs with 
core collapse of massive Wolf-Rayet stars. The possibility of
GRB to be a transient phenomenon in the early history of galactic star
formation related to evolution of very massive metal-free stars 
is briefly outlined.   

\end{abstract}

\vspace*{0.5cm}

\section*{Introduction and history of GRB studies}

Cosmic gamma-ray bursts (GRB) continues to be one of the hottest points in
modern astrophysics in spite of a significant progress achieved in this
field over the last few years. The state of the art in this field of
high-energy astrophysics has been reviewed many times (for example,
\cite{Luchkov_ea96, Postnov99, Piran99, Lamb00} and references
therein), but the new results appear so rapidly that any review can hardly
take into full account all fresh facts and ideas. We start with short
description of basic observational facts about GRBs and associated
astronomical phenomena, referring the reader to the above mentioned reviews
for more detail. 

Cosmic GRBs were spuriously discovered in the end of 60s as
short (1-100 s), highly variable (on time scales down to 1 ms) bursts of
hard gamma-rays ($E>10-30$ keV up to Gev and sometimes TeV hard tails) with
non-thermal spectra (typically approximated by two power laws \cite{Band_ea}
with a maximum energy release at around several hundred keV), which appear
randomly over sky. No spectral lines have been 
firmly found in the GRB spectra.

The typical fluence (i.e. the energy integrated over the duration of the
burst) spans from $S_{min}\sim 10^{-7}$ ergs\footnote{The lower limit is
determined by the detector sensitivity} for the faintest bursts up to
$S_{max}\sim 10^{-3}$ ergs for the brightest ones. The early studies
(roughly, up to the launch of the specialized BATSE all-sky gamma-ray
monitor onboard Compton Gamma-Ray Observatory in 1991) established the
isotropic distribution of GRBs over the sky, which was later reliably
confirmed by BATSE, and a great variety of properties of individual bursts.
A poor localization of GRB with typical error boxes of several degrees at
that time prevented GRB from being reliably identified with known
astrophysical sources. Since no way has been known how to determine 
the distance to the source of gamma-ray emission, even such an important
characteristic as the total energy emitted was unknown at that time, which
causes a lot of speculation about GRB models from the local plasma events
inside the solar system up to cosmological sources.

The main important result obtained by BATSE (1991-2000) was the
understanding that GRBs are not distributed uniformly in space. Test $\log
N$--$\log S$ (cumulative distribution of sources with fluxes (fluences)
higher than a given value) constructed using BATSE observations reliably
showed a significant deviation of faint bursts from the $N\propto S^{-3/2}$
law expected in the Euclidean space, which strongly indicated a cosmological
origin of GRBs. The last clue for non-galactic origin of GRBs came from
other than gamma-ray observations, initiated by the launch of Dutch-Italian
satellite BeppoSax. Observations of X-ray
\cite{Costa_ea97}, optical \cite{Groot_ea97a} and radio
\cite{Frail_ea97} afterglow of GRBs put the end to the isolation of GRB
studies as a specific high-energy astrophysical problem 
from other astronomical observations. A high-accuracy (several arc minutes) 
localization
of GRB sites became possible, opening up the possibility to search for
known astronomical objects within error boxes of GRBs with optical
afterglows. These searches immediately established the connection of
GRBs with remote host galaxies \cite{Sahu_ea97}.

By the end of September 2000, of 84 GRBs with error boxes better than a few
arc minutes, X-ray afterglows were observed from 35, optical from 21 and
radio from 15 GRBs (see \cite{Greiner00}). Redshift has been measured for
more than 10 GRBs (afterglows or hosts), which allowed to recover the
luminosities and total energy emitted in the burst. The typical luminosities
(assuming isotropy of gamma-ray emission) fall within a broad range from
$\sim 5\times 10^{51}$ ergs to $\sim 2\times 10^{54}$ ergs. An apparently
separate case is GRB980425 associated with a peculiar type Ic
supernova\footnote{Type Ic SNe includes those displaying no hydrogen and
helium lines in their early spectra; their most likely progenitors are bare
CO cores of massive stars} 1998bw in a nearby ESO 184-G82 ($z=0.0085$)
\cite{Galama_ea98}. Having quite typical GRB properties (spectrum, duration,  
light curve), the energy released in this bursts is as small as $\Delta
E_\gamma\approx 10^{48}$ ergs, which is by four orders of magnitude smaller
than the typical value $5\times 10^{51}$ ergs. The peculiarity of this
supernova is that an unusually high kinetic energy ($\sim 6\times
10^{51}-10^{52}$ ergs) is required to model the observed light curve
\cite{Iwamoto_ea98, Woosley_ea99}. 

Cosmological distances to GRBs immediately posed an important physical
problem for GRB models, the so-called compactness problem, which importance
was realized much earlier 
\cite{Guilbert_ea83}. The short time-scale of the observed
variability ($\sim 10^{-2}$ s) implies a small volume of the emission, and
the large energy emitted implies an enormous photon optical depth
$\tau_{\gamma\gamma}>10^{12}$ for electron-positron pair creation
$\gamma+\gamma\to e^++e^-$. Such a huge optical depth would lead to
thermalisation of photons which is in conflict with the observed non-thermal
spectra of GRBs. The problem is elegantly resolved assuming a relativistic
velocity of the emission region expansion (see \cite{Blinnikov00} for
detailed discussion and full reference), and it is now widely believed that
the GRB phenomenon itself is related to relativistic shock waves initiated
by a photon-lepton fireball, expanding with an initial Lorentz-factor of
$\Gamma\sim 200$ in the interstellar medium \cite{Rees&Meszaros92} (see
details and full references in \cite{Piran99}). 

In the framework of this model, the GRB phenomenon is due
to some (reliably unknown) energy release (explosion) in the form of photons
and leptons with a small ($\sim 10^{-5} M_\odot$) load of baryons. During
expansion, the initial thermal energy of the fireball is converted to the
kinetic energy of relativistic blast wave that grubs the surrounding matter
and brakes down thus converting its kinetic energy to the energy of
relativistic particles (electrons) at the blast wave front. Thermal energy
of relativistic electrons are radiated by synchrotron emission in the
magnetic field giving rise to the observed X-ray, optical, and radio
afterglows. The so-called "internal shock wave" model assumes that the GRB
itself is generated during interaction of individual shock waves with each
other, and the waves themselves appear during the initial energy release by
an unknown "central engine". Despite the simplicity and elegance of this
model, it is still far away from fully adequate description of the observed
properties of GRBs (see \cite{Usov99} for an alternative model and
criticism).

The requirements to the central engine of GRBs are mainly reduced to the
following: (1) The ability to release the electromagnetic energy $\sim
10^{52}$ ergs during 10-100 s (the typical duration of "long" GRBs, only for
which X-ray and optical observations are possible; a separate group of
short single-peak GRBs is much less studied, apart from the fact 
that they are isotropically
distributed over the sky, and well may be another phenomenon) and (2)
The event rate is on average about one burst per typical galaxy (assuming
isotropy of the emission and homogeneity of galaxies). Clearly, beaming of
gamma-ray emission will decrease the energy emitted and increase by the same
amount the event rate.

These requirements are met (with different degree of accuracy) by
several classes of astrophysical sources. Mostly cited models include:

(1). Coalescence of binary neutron stars and/or black holes,
originally suggested by Blinnikov et. al. in 1974  
\cite{Blinnikov_ea84}). The fireball is generated by 
neutrino-antineutrino annihilation copiously produced 
during the coalescence.  

(2). Hypernova model suggested by Paczy\'nski \cite{Paczynski97},
in which the energy is extracted from a rapidly rotating 
massive  ($\sim 10 M_\odot$) black hole surrounded
with a disk threaded by a superstrong magnetic field of 
$\sim 10^{15}$ G by the Blandford-Znajek mechanism
\cite{Blandford&Znajek76}. A closely related model 
by Woosley \cite{Woosley93, Macfadyen&Woosley99}
involves the formation of an accretion disk around a 
massive rotating black hole during late stages of 
the core collapse of a massive star; in this model, 
a narrow jet is produced inside the star and punches
through the stellar envelope reaching very high Lorentz-factors.

(3). Electromagnetic model by V.Usov 
\cite{Usov92}, in which the energy comes from the rapid rotation of 
a young neutron star (millisecond pulsar) with a very strong
magnetic field. 

(4). Recently, S.S.Gershtein proposed a model of
GRB during core collapse of a non-rotating Wolf-Rayet star
\cite{Gershtein00}, in which the internal shocks are
created due to the collapse non-stationarity and energy 
is brought away by electron-positron plasma. 

Here we focus on modern astronomical observations which mostly constraint
the nature of cosmic GRBs. In particular, we shall consider
energy emitted and the 
luminosity function of GRBs; host galaxies of GRBs and their
properties; association of GRBs with strong star formation 
regions. We also will discuss the observed event rate of GRBs
and its implications. Many other properties of GRB themselves 
are discussed in more detail elsewhere in this volume \cite{Stern00}.
     
\section*{Energetics of cosmic GRBs}

\begin{table}
\caption{GRBs with known energetics}
\label{GRB}
\begin{tabular}{lccccccc}
\hline
\hline            
GRB            & z     & $d_l^\dag$,  $10^{28}$& $F_\gamma$, $10^{-5}$ & Ref& $\Delta E$, $10^{53}$ & $F_p^\ddag$  & $L_p$, $10^{58}$ \\
               &       & cm& erg/cm$^2$           &     & erg& ph/s/cm$^2$ & ph/s \\
               &       &             &(10-2000 keV)         &     &             & (50-300
keV) \\
\hline
000926         & 2.066:& 5.81     &2.2  &\protect{\cite{JPUFynbo_ea00GCN807}}& 3.04
\\
000418$^{a)}$  & 1.118 & 2.73     &1.3  & \protect{\cite{JBloom_eaGCN661}}       &$\sim 0.6$ &  -    \\
000301C$^{b)}$& 2.03  & 5.69     &$>0.05$ & \protect{\cite{SMCastro_eaGCN605}}     &$\sim 0.07$& $\sim 5$ & 6.7    \\
991208$^{c)}$& 0.706 & 1.55     &10   & \protect{\cite{SGDjorgovski_eaGCN481}} &$\sim 1.8 $&  -     \\
990712         & 0.430 & 0.85     &-      \\
990510	       & 1.619 & 4.30     &2.26 & \protect{\cite{Kumar&Piran00}}         &2.0        &8.16& 7.3    \\
990123	       & 1.6   & 4.25     &26.8 & \protect{\cite{Kumar&Piran00}}         &23         &16.4& 14    \\    
980703         & 0.967 & 2.28     &2.26 & \protect{\cite{Kumar&Piran00}}         &0.75       &2.6& 0.86  \\
980613$^{d)}$  & 1.096 & 2.66     &0.17 & \protect{\cite{Briggs_ea99}}
&0.072         &0.63& 0.27  \\
971214         & 3.412 & 10.6     &0.944& \protect{\cite{Kumar&Piran00}}         &3.0        &2.3 & 7.4  \\
970828	       & 0.958 & 2.25     &9.6  & \protect{\cite{Kumar&Piran00}}         &3.1        & - \\
970508	       & 0.835 & 1.99     &0.317& \protect{\cite{Kumar&Piran00}}         &0.08       &1.2& 0.29  \\
970228$^{d)}$  & 0.695 & 1.52     &$\sim 0.2$& \protect{\cite{Briggs_ea99}}
&0.034          &3.5& 0.60  \\
\hline
980425         & 0.0085 & 0.013   & 0.32 & \protect{\cite{Galama_ea98}} & $7\times
10^{-6}$\\ 
\hline
\end{tabular}
\medskip

Notes:

$\dag$ Flat Universe, $\Omega_m=0.3, \Omega_\Lambda=0.7, H_0=60$~km/s/Mpc 

$\ddag$ Peak fluxes from \protect{\cite{Lamb00}}  

a) Energy fluence in 25-100 keV;

b) Peak flux $F_p=3.7$ Crab, no fluence published, 
single-peak profile, duration 10 s; other indirect estimations of
total energy in gamma-rays see in \protect{\cite{Sagar_ea}} 

c) Energy fluence for $E>25$ keV

d) Energy fluence for $E>20$ keV
\end{table}

As mentioned in the Introduction, the energy released in GRB
with known redshifts is spread over a broad range. It is rather difficult
to precisely recover this energy even for 
bursts with known redshifts due to background flux fluctuations, 
varying spectrum etc.  For homogeneous set of BATSE bursts 
\cite{Kumar&Piran00} energies are calculated in Table \ref{GRB}. 
Additionally, we included some new non-BATSE bursts with measured redshifts
GRB991208, GRB000301C, and GRB000418. Photometric distances
were determined using a flat Universe with $\Omega_m=0.3,
\Omega_\Lambda=0.7, H_0=60$~km/s/Mpc. The Table contains also 
BATSE fluences (50-300 keV) and peak luminosities $L_p$ phot/s
(from \cite{Lamb00}). The latter quantity is less affected by 
selection effects and characterizes the internal energy release.

\begin{figure}
\centerline{
\hss
\epsfysize=7cm
\epsfbox[100 110 500 650]{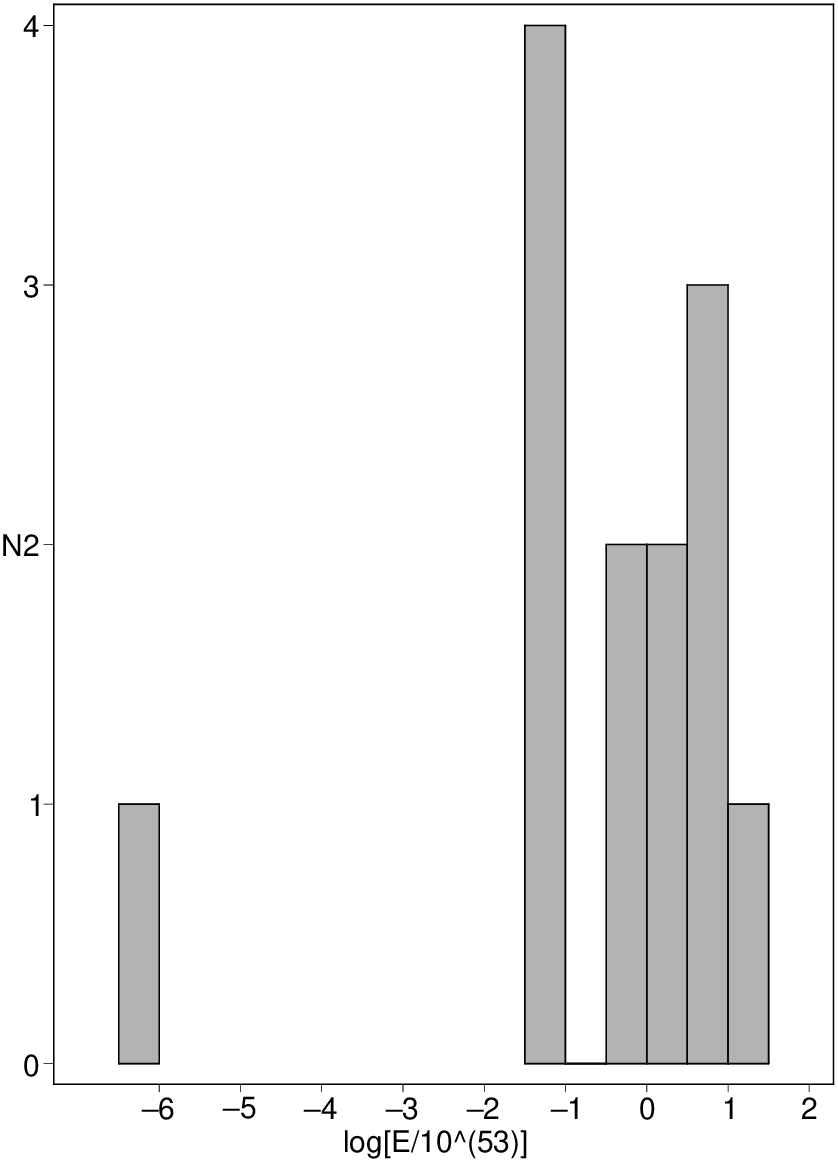}
\hss
\epsfysize=7cm
\epsfbox[100 110 500 650]{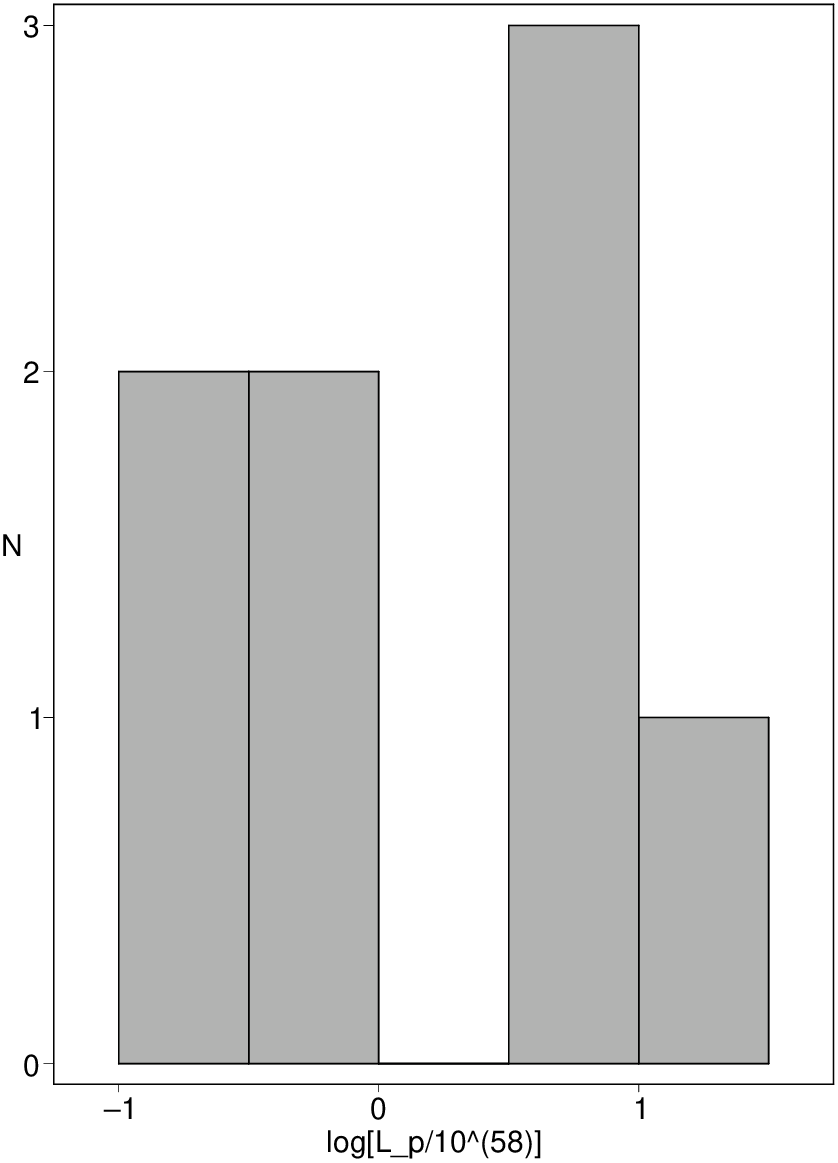}
\hss
}
\caption{Total GRB energy release $\Delta E$ 
(in $10^{53}$ erg/s) of GRBs with known redshifts 
from Table \protect\ref{GRB} }
\label{energy}
\caption{Peak luminosities $L_p$ (in 
$10^{58}$ phot/s) 
of GRBs from Table \protect\ref{GRB}}
\label{flux}
\end{figure}

As seen from Table
\ref{GRB} and Fig. \ref{energy}, \ref{flux},
the observed GRB energetics spans from  
$\approx 7\times 10^{51}$ ergs to  $\approx 2\times  10^{54}$ ergs. 
This observational fact is usually explained by 
a broad luminosity function of GRBs
(see \cite{Loredo&Wasserman98}, \cite{Schmidt00}),
although one can construct a self-consistent model with
a universal energy release of 
$E_0\sim 5\times 10^{51}$ ergs and a complex beam shape  
\cite{LPP00}. Note that adding GRB980425 either evidences
for a bimodality of GRB energy distribution \cite{Postnov&Cherepashchuk00} 
or a extremely broad luminosity function (more than 5 orders of magnitude!).

\section*{GRB phenomenon connection with very massive stars}

Dedicated optical observations of the identified host galaxies 
of GRBs carried out by the largest ground-based telescopes and
the Hubble Space Telescope \cite{HST00} revealed that 
all these galaxies are somewhat peculiar, either 
by morphology (if the galaxy is sufficiently resolved)
or by color. All evidences for an enhanced star formation rate
inside the GRB sites, sometimes by an order of magnitude higher
than in our Galaxy \cite{Vreeswijk_ea00}. Independently, 
an analysis of known X-ray and optical afterglows of GRBs
showed a large column densities
of $10^{22}-10^{23}$~cm$^{-2}$ toward GRBs 
\cite{Galama&Wijers00}, typical for giant
molecular clouds. This strongly suggests that at least these
GRBs are associated with evolution of massive stars, and makes
other GRB progenitors (such as double neutron star 
coalescences) less likely.  Here one should be cautious 
with final conclusions and wait for more statistics, 
which hopefully will be available with upcoming 
launch of specialized satellites HETE-2 and SWIFT. 
For example, double neutron star mergers could show 
less pronounced afterglows because of smaller ambient
densities, etc. Moreover, for double neutron star merging
scenario there is a definitive prediction \cite{Panchenko_ea99}
that can be checked by observations -- such events should
take place both in spiral and elliptical galaxies (presently having
practically no star formation), with the 
fraction of elliptical hosts increasing up to 30-40\% with
redshift, while the hypernova scenario always requires a 
tight connection with star forming regions. So the discovery 
of an elliptical host without pronounced star formation would 
evidence for non-universality of the hypernova scenario. 

Here, however, we will concentrate on implications of recent observations on
possible GRB progenitors. The evolution of massive stars takes a relatively
short time (a few million years) and ends up with core collapse, which 
is associated with supernova explosion of type II or type Ib/c. In the
latter case the core collapse occurs in a star deprived of its hydrogen
envelope (so-called Wolf-Rayet (WR) stars). The possible relation of GRBs
with WR stars were discussed in
\cite{Postnov&Cherepashchuk00}. The arguments 
favoring such an association are as follows \cite{Postnov&Cherepashchuk00}.

1) {\it Energetics}. The observed GRB energy release 
$\Delta E\simeq 10^{51}-10^{54}$ erg is roughly comparable
with the wide range of CO cores of WR stars before collapse, 
$M_{CO}^f\simeq 2-40 M_\odot$. The total energy released
in collapse is $E_G\sim GM_c^2/R_c$, where $M_c$ in the compact remnant
mass, $R_c$ its radius. During black hole formation in such collapses
without mass ejection the available energy can reach
$10^{53}-10^{56}$ ergs for the observed mass range of
compact remnants ($M_{NS}=1.35\pm 0.15M_\odot, M_{BH}=5\div 15 M_\odot$
\cite{Cherepashchuk01}, i.e. conversion of 1\% of the available
energy into kinetic energy of shocks with subsequent radiation would be
sufficient to explain the broad luminosity function. Note that during
collapse into black hole without mass ejection
$R_c\propto M_c\propto M_{CO}$ and energy range is proportional 
to the core mass range, which seems to really take place. 
Note that possible gamma-ray beaming decreases the total energy
release but does not narrow the luminosity function.

2) {\it Bimodality of GRB energy distribution and stellar remnant mass
distribution}. GRBs with low energetics associated with peculiar supernovae
type Ic (such as GRB980425) can be explained by collapses of bare CO cores
of massive stars with significant rotation which causes most envelope to be
ejected and neutron star to be formed, while collapses of slower rotating
cores do not accompanied by a significant envelope ejection and lead to
black hole formation. In the latter case an energetic GRB can be generated
with energy proportional to the pre-collapse core mass.
  
3) {\it Association of GRBs 
with star forming regions}. It is natural in all models 
invoking massive star evolution. 

4) {\it A diversity of the observed afterglows}.
As was first noted in \cite{Blinnikov&Postnov97},
a GRB occurring in a binary system can induce different optical
phenomena due to illumination of the companion's atmosphere
by hard X-ray and gamma-radiation. This should add some light
to the "pure" power-law afterglow from relativistic blast wave
thus producing a great variety of the observed light curves.  
These effects can occur in a time interval   
$\Delta t_{opt}>D/c$ after the burst 
($D$ is the distance to the optical star from GRB, $c$ is the 
speed of light). 
Deviations are indeed observed
in some bursts (for example, in GRB980326 afterglow 
three weeks after the burst \cite{JSBloom_ea99}).

Astronomical observations indicate \cite{Cherepashchuk01, Cherepashchuk98,
Cherepashchuk2000} that about 50\% of all WR stars in our Galaxy
can be in binaries with O-star or A-M-star.  
For example, for WR+O system V444 Cyg with an orbital period 
of $P=4^d.2$ we have $D\approx 40 R\odot$ and the time delay 
$\Delta t_{opt}\approx 100$~s, and for 
parameters of WR+O binary system 
CV Ser $\Delta t_{opt}\approx 300$~s.
Note that an extremely bright optical emission ($V\simeq 9^m$) 
was observed in the famous burst GRB990123 
only 50 s after the burst beginning \cite{Akerlof_ea99}.

\begin{figure}
\centerline{
\epsfysize=10cm
\epsfbox[50 50 700 700]{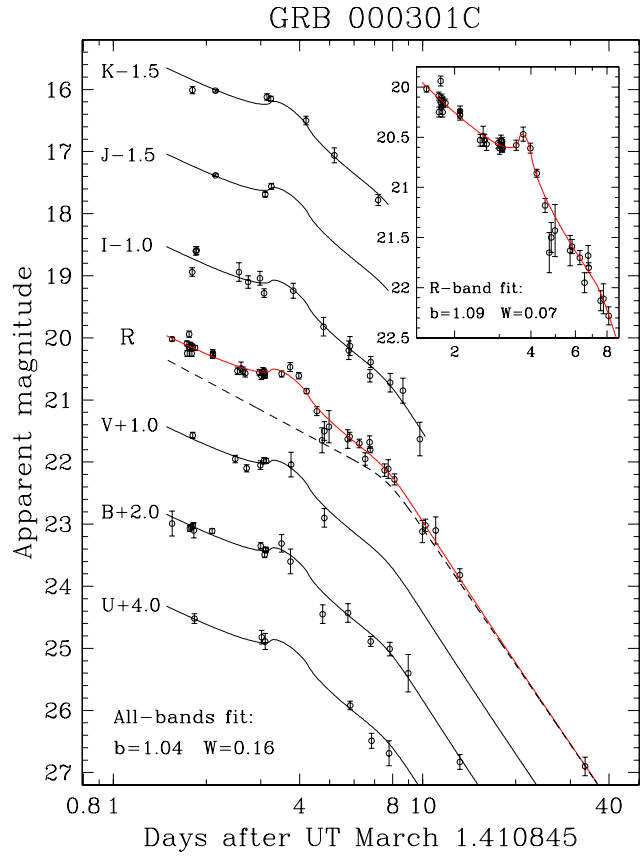}
}
\caption{Observed multicolor optical-IR afterglow
of GRB000301c (adapted from \protect\cite{Garnavich_ea00}).
Peaks separated by several days are clearly seen in all bands.}
\label{000301c}
\end{figure}

Another example is a peculiar shape of achromatic optical afterglow light
curve observed in GRB000301c
\cite{Masetti_ea00, Sagar_ea00} (see Fig. \ref{000301c}. 
The observed several peaks  separated by 2-3 days 
days can be a manifestation of an 
orbital period in the underlied binary system, for example, through the
binary-period shaped mass loss before collapse. An alternative explanation
by a microlensing event \cite{Garnavich_ea00} seems less probable. 
Orbital periods of order of several days perfectly fit 
the observed period range
$1^d.6\div 2900^d$ in WR+O binary systems (see Table 2 in
\cite{Postnov&Cherepashchuk00}).

These arguments favor the GRB-WR stars association, but
there is a general requirement which should be met by all 
viable GRB models. The point is that GRB phenomenon should be an
extremely rare astronomical event. 
  
\section*{Event rate problem}

Selection effects make it difficult 
to reliably estimate the event rate of GRBs
using current observations. A careful analysis 
of non-triggered BATSE GRBs carried out by 
B.E.Stern et al. \cite{Stern_ea00} shows that 
the total number of BATSE bursts for a threshold 
flux level of  $F_{tr}=0.1$ ph/cm$^2$ is 
1200-1300 per year. This implies the GRB event rate per
unit comoving volume 3-4 times higher than was previously
calculated using BATSE data with $F_{tr}=0.5$ ph/cm$^2$
\cite{Schmidt00}  and is 
$$
{\cal R}_{GRB}\sim 10^{-9} \hbox{GRB/yr/Mpc}^3\,,
$$
i.e about $10^{-6}-10^{-7}$ per year in the average galaxy 
with a mass of $10^{11} M_\odot$. This is by several orders of magnitude
lower than the total rate of core collapses associated with 
SN II and Ibc
(${\cal R}_{SNIbc}\sim 3 \times 10^{-5}$/ÇÏÄ/íÐË$^3$,
\cite{Cappellaro_ea97}). This discrepancy is usually eliminated 
by introducing a beaming of gamma-ray emission   
(e.g. \cite{Lamb00}). It is not excluded that not each SN Ibc
is associated with GRB for internal reasons. 

It is straightforward to estimate the mean formation rate 
of WR stars in the Galaxy. Let the galactic star formation rate is
constant. The birth rate of solar-type stars in the Galaxy 
${\cal R_\odot}$ is about 1 star per year, and the 
total number of such stars in the Galaxy
is $N_\odot\simeq 10^{11}$. The galactic number of WR
stars is $10^{3}\div 5\times 10^{4}$ (assume $N_{WR}=2\times 10^3$).
Since the lifetime of a solar-type star is 
$\Delta t_\odot\simeq 10^{10}$ years and the mean lifetime
of a WR star  is $\Delta t_{WR}\simeq 5\times 10^5$ years,
the WR birth rate will be  
\begin{equation}
{\cal R}_{WR} = {\cal} R_\odot\frac{N_{WR}}{N_\odot}
\frac{\Delta t_\odot}{\Delta t_{WR}}\sim \frac{1}{1000} \hbox{yr}^{-1}\,.
\end{equation}

It is seen that the WR birthrate by a factor of 1000 exceeds that of GRBs,
and this discrepancy should be explained. This can be done either by
postulating generically thin jets or, admitting quasi-spherically symmetric
emission, by assuming the existence of some "hidden" collapse parameters
(rotation, magnetic field, etc.) The important role of such parameters for
the outcome of collapse was also suggested in \cite{Ergma&vdH98} from an
independent analysis of black hole formation in binaries. In the hypernova
scenario by Paczy\'nski \cite{Paczynski97} the rarity of GRB phenomenon is
explained by requirement of an extremely high magnetic field during core
collapse of a massive star into a 10 $M_\odot$ black hole.

In contrast, in the model of coalescing neutron star/black hole binaries
(which is currently less favored by association of all observed GRB hosts
with strong star forming regions, see above) the event rates varies from
$\sim 10^{-4}$ to $\sim 10^{-6}$ per year depending on the binary 
evolution parameters \cite{LPP97}, which is marginally consistent
with the observed GRB rate and the event rate problem 
is not very strong. 

\section*{GRBs as a transient galactic phenomenon}

There is another possibility of explanation of the observed association of
cosmic GRBs with star formating regions at high redshifts and their extreme
rarity. GRBs may be a transient galactic phenomenon occurring at the early
stages of galactic evolution, like quasars and AGNs. It is established now
\cite{Madau_ea96} that at high redshifts $z\sim 1-2$ a violent epoch of star
formation in young galaxies occurred. It is also known that a lot of cold
matter were bound in giant proto-galactic clouds at redshifts $z>2$, which
are observed as "Lyman-alpha forest" of absorption lines in quasar spectra.
The formation of very massive stars 100-500 $M_\odot$ which final 
collapse into massive
black holes took place at that epoch. Such 
massive star can not form from matter enriched with metals
because of pulsational instabilities (see \cite{Baraffe_ea00aph/009410}
and references therein). 
At low metallicity at the epoch of violent
star formation, however, they could have formed. The 
possibility of energetic GRBs from collapses of such massive stars
was studied in \cite{Fryer_ea00aph/0007176} with negative conclusion 
about their ability to produce an energetic GRB. But we note here that 
physical processes in such stars are still far from full 
understanding and potentially such stars could be 
GRB progenitors. Note that the weakness of GRB980425 in a nearby
galaxy can be a natural consequence of smaller upper masses of  
stars in regions of violent star formation at 
the present epoch.

\section*{Conclusion}

We considered some important constraints that modern astronomical
observations provide for GRB models. First of all, this is the apparently
broad (possibly bimodal) energy release in GRBs with known redshifts. Such
a distribution can be explained either by assuming an intrinsically
broad luminosity function, or a wide spread of beaming angles, or else 
by proportionality of this energy to the mass of collapsing CO cores.

Next important point is the observed GRB event rate, roughly one per million
years per average galaxy, which is by 3 orders of magnitude higher than the
birthrate of massive WR stars in our Galaxy. If GRBs are associated with the
evolution of massive stars, as the growing evidence from observations of the
identified GRB host galaxies suggests, this discrepancy can be explained
either by assuming beaming of gamma-rays (jets) or random outcome of the
core collapse.

At last, the GRB phenomenon can be associated with evolution of very massive 
metal-free stars, which could have formed at the early stage of galactic 
evolution during the epoch of violent star formation. If so, GRBs can be
a transient phenomenon in the evolution of galaxies, and only weak
GRBs can be produced by core collapses at present epoch.

As we noted in \cite{Blinnikov&Postnov97}, 
the huge energy released during a GRB in a galaxy can not 
be passed without imprints in the surrounding interstellar
medium, such as huge caverns in the interstellar gas or 
enhanced  star
formation induced by the explosion.  
Stellar studies of LMC and other nearby galaxies 
show the presence of giant stellar arcs and rings 
which may be relics of giant GRB explosions in these galaxies
(see \cite{Efremov00} and references therein). This demonstrates
the potential important role of careful investigations
of GRB-related astronomical effects.

There are other, more exotic models of GRBs, which satisfy these
requirements. We mention here an interesting model by S.Blinnikov
\cite{Blinnikov00}, who suggested that GRBs can be produced by 
violent events in the hypothetical mirror world (such as coalescence
of mirror binary neutron stars). At present, we cannot choose between
many possibilities. But with upcoming new experiments HETE-2 and SWIFT, 
a lot of new discoveries of GRBs and their simultaneous observations
at softer wavelength will become possible, which undoubtedly will
help discriminate between many models of energetic GRBs. 

\vskip\baselineskip 
{\it Acknowledgements} The author acknowledges the staff of Bogolyubov
Theoretical Physics Laboratory of JINR for hospitality and support, and
A.M.Cherepashchuk, M.E.Prokhorov, S.I.Blinnikov, Yu.N.Efremov for
discussions. The work was partially supported by Russian Fund for Basic
Research through Grants 99-02-16205 and  00-02-17164.

\end{document}